\documentstyle[prl,epsfig,aps]{revtex}
\begin{document}
\draft
\newcommand{\be}{\begin{equation}}
\newcommand{\ee}{\end{equation}}
\newcommand{\bea}{\begin{eqnarray}}
\newcommand{\eea}{\end{eqnarray}}

\title{Quantizing Billiards with Arbitrary Trajectories}
\author{Debabrata Biswas\thanks{email: biswas@alf.nbi.dk}\thanks{to 
appear in the Proceedings of the Int. Conf. on Non-Linear Dynamics and Computational
Physics, {\it Physical Research Laboratory, Ahmedabad, 1997}}}
\address{Center for Chaos and Turbulence Studies,
Niels Bohr Institute \\
Blegdamsvej 17,  
Copenhagen $\O$, Denmark \\ \& \\
Theoretical Physics Division \\
Bhabha Atomic Research Centre \\
Mumbai 400 085}
\maketitle
\begin{abstract}

The structure of the semiclassical trace formula can be used
to construct a quasi-classical evolution operator whose
spectrum has a one-to-one correspondence with the semiclassical
quantum spectrum. We illustrate this for marginally unstable
integrable and non-integrable billiards and demonstrate its 
utility by quantizing them using arbitrary non-periodic 
trajectories.

\end{abstract}

\vskip 0.25 in

\nopagebreak

\par Periodic orbits form the skeleton around which generic
classical motion is built, and not surprisingly, many dynamical
quantities of interest can be computed quite accurately 
in terms of these invariant trajectories \cite{PC}.
They also form the basis of modern semiclassical
quantization schemes which express the trace of  
the energy dependent quantum propagator 
in terms of lengths, stabilities and focal points of periodic
orbits \cite{MC}. Thus resonances in open systems and
the energy eigenspectra of closed systems can be determined
semiclassically even when the dynamics is chaotic.
  
\par A similar relationship also exists for the trace of the 
classical propagator, $L^t$, though this is known explicitly
only when the system is hyperbolic \cite{PCBE}. Here $L^t$
governs the evolution of densities  

\be
L^t{\circ}\phi({\bf x}) = \int {\bf dy}\; \delta({\bf x} -
{\bf f}^t({\bf y}))\;\phi({\bf y}) \label{eq:def1} 
\ee

\noindent
where {\bf x} = ({\bf q,p}) and ${\bf f}^t$ refers to the flow in
the full phase space. We denote by $\Lambda_n(t)$ the eigenvalue
corresponding to an eigenfunction $\phi_n({\bf x})$ \cite{notL2}
such that
$L^t{\circ}\phi_n({\bf x}) = \Lambda_n(t) \phi_n({\bf x})$.
It then follows that the autocorrelation of any analytic function,
$A(x(t))$ can be expressed as 
$\sum_n a_n \Lambda_n(t)$ where the coefficients $\{a_n\}$
are determined by $A({\bf x})$ and the eigenfunctions 
$\{\phi_n({\bf x})\}$.
The classical spectrum, $\{\Lambda_n(t)\}$, is thus crucial
to understanding the evolution of correlations and the first 
step towards determining this is to evaluate the trace
of $L^t$. For hyperbolic systems \cite{PCBE} 

\be
{\rm Tr}~L^t = \sum_n \Lambda_n(t) = \sum_p \sum_{r=1}^{\infty}
{T_p \delta(t-rT_p)
\over \left | \det({\bf 1} - {{\bf J}_p}^r \right | } \label{eq:hbolic}
\ee

\noindent
where the summation over p refers to all primitive periodic 
orbits, $T_p$ is the time period and ${\bf J}_p$ the Jacobian 
matrix evaluated on the orbit.
Note that the semi-group property, $L^{t_1}\circ L^{t_2}
= L^{t_1 + t_2}$, for continuous time implies that
the eigenvalues $\Lambda_n(t)$ are of the form $e^{\lambda_n t}$.
The poles in the Laplace transform of ${\rm Tr}~L^t$ thus
occur at {$\lambda_n$} much in the same way as the trace
of the energy-dependent quantum propagator has poles at the 
eigenenergies, $\{E_n\}$, of the system.
There are important differences though between the traces
of the quantum and classical evolution operators. The
quantum case involves the square root
of $\left | \det({\bf 1} - {{\bf J}_p}^r \right |$ and
in addition has Maslov phases so that 
the classical spectrum in general bears no 
relationship to the quantum spectrum \cite{exception}.

\par Our intention here is to construct  
quasi-classical propagators similar to Eq.~(\ref{eq:def1}),
but based on the quantum trace formula so that $\{\lambda_n\}$
is directly related to $\{E_n\}$. This can be used to our
advantage by evolving classical functions and extracting
the semiclassical spectrum from the peak positions in its 
power spectrum. Importantly, the recipe does not involve
periodic orbits which often have a highly non-trivial
topological organisation due to which a systematic 
computation is difficult. To keep matters simple however,
we shall carry out this exercise for both integrable
and non-integrable systems that are marginally
unstable. Compared to chaotic systems, these are
simpler to handle by far but nevertheless serve to illustrate
the essential idea.
 
\par Quasi-classical propagators have been studied
before by Cvitanovi{\'c} and Vattay \cite{vattay1,PC} though 
from a different standpoint. They consider an extended
dynamical space to evolve the quasi-classical wavefunction
and ultimately derive a trace formula that is distinct
from the usual Gutzwiller trace formula for chaotic systems, 
and, whose corresponding Fredholm determinant is entire.
Thus, they do not consider the question of quantization
using arbitrary trajectories. A resemblance to the 
method employed here can however be found in the 
approach of Heller and Tomsovic \cite{heller} in that
they work in the time domain and use the fourier
transform of the correlation function to determine the
spectrum, $\{E_n\}$. There are basic differences
though, primarily concerning the nature of the 
evolution. While we use here the classical propagator
and its quasi-classical adaptation, Heller and
Tomsovic \cite{heller} use a semiclassical approximation of the
quantum propagator. As a consequence, the quantities
that need to be computed differ.  
  

\par We shall, without much loss of generality, confine
ourselves to billiards \cite{dbprep}. The ones that are
marginally unstable include the integrable circle billiard
and rational polygonal billiards. An important 
characteristic of these systems is the existence of two
independent constants of motion due to which the invariant
surface is two dimensional. In case of integrable motion,
this is topologically equivalent to a torus while in the
general case, the invariant surface has the topology of
a sphere with $g$ holes. Examples of integrable billiards
are few and generic rational polygonal billiards are
pseudo-integrable \cite{PJ} with $g > 2$.
   
\par Before considering the question of quasi-classical
quantization, we first introduce the appropriate 
{\it classical} evolution operator. For
integrable systems, this is easily defined as 

\be
L^t{\circ}\phi(\theta_1,\theta_2) =
\int d\theta_{1}'d\theta_{2}'\;
\delta (\theta_1 - \theta_{1}'^t)\delta (\theta_2 - \theta_{2}'^t)
\;\phi(\theta_1',\theta_2') \label{eq:prop}
\ee

\noindent
where $\theta_1$ and $\theta_2$ are the angular coordinates on the
torus and evolve in time as $\theta_i^t = \omega_i (I_1,I_2)t +
\theta_i$ with $\omega_i = \partial H(I_1,I_2)/\partial I_i $ and
$I_i = {1\over 2\pi}\oint_ {\Gamma_i} {\bf p.dq} $.
Here $\Gamma _i, i = 1,2$ refer to the
two irreducible circuits on the torus and ${\bf p}$ is the momentum
conjugate to the coordinate ${\bf q}$.

\par  It is easy to see that the eigenfunctions 
$\{ \phi_n(\theta_1,\theta_2)\}$ are such that
$\phi_n(\theta_1^t,\theta_2^t) = \Lambda_n(t) \phi_n(\theta_1,\theta_2)$
where $\Lambda_n(t) = e^{i\alpha_n t}$. On demanding that
$\phi_n(\theta_1,\theta_2)$
be a single valued function of $(\theta_1,\theta_2)$, it follows
that $\phi_{\bf n}(\theta_1,\theta_2) = e^{i(n_1\theta_1 + n_2\theta_2)}$
where ${\bf n} = (n_1,n_2)$  is a point on the integer lattice.
Thus the eigenvalue,
$\Lambda_{\bf n}(t) = {\rm exp}\{it(n_1\omega_1 + n_2\omega_2)\}$. 
We shall evaluate this explicitly for the rectangular billiard
while discussing the 
trace of $L^t$ and we now introduce the appropriate evolution
operator that applies also to rational polygonal billiards 
which are generically non-integrable.

\par For both integrable and pseudo-integrable polygonal 
billiards,
the dynamics in phase space can be viewed in a singly connected region by
executing $2g$ cuts in the invariant surface and identifying 
edges appropriately. 

\par
\vbox{
\begin{figure}[hbt]
\centerline{\hspace{0.85in}
\epsfig{figure=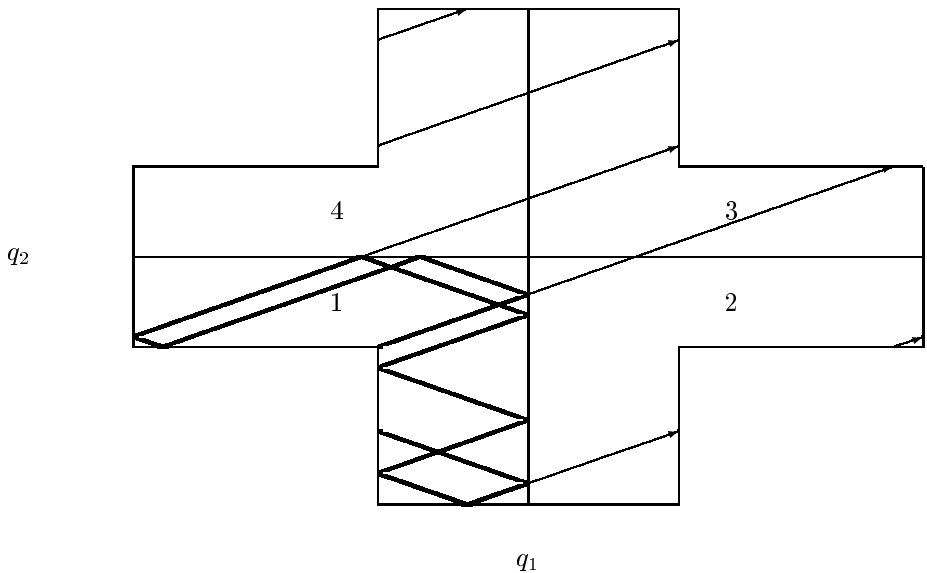,height=3.5in}}
\caption{The singly connected region for an L-shaped billiard consists
of four copies with edges appropriately identified. A trajectory originating
near the $3\pi/2$ vertex in $1$ is plotted in configuration space using bold
lines and the corresponding unfolded trajectory is also shown.
The latter consists of parallel segments and the trajectory can be parametrized
by the angle $\varphi$ that it makes for example with the $q_1$ axis.}
\end{figure}
}

\noindent
At a given energy, the motion is
parametrised by the angle, $\varphi$, that a trajectory
makes with respect to one of the edges. As a trivial example,
consider the rectangular billiard. 
The singly connected region is a larger rectangle consisting
of four copies corresponding
to the four directions that a trajectory can have and these
can be glued appropriately to form a torus \cite{keller}. 
As a non-trivial
example (see Fig~1), consider the L-shaped billiard which is
pseudo-integrable with its invariant surface having, $g=2$.
Alternately, the surface can be represented by a singly connected
region in the plane and consists of four copies
corresponding to the four possible directions an orbit can
have and these are glued appropriately.
A trajectory in phase space thus consists of parallel segments at an 
angle $\varphi$ measured for example with respect to one of the sides.
It will be useful to note at this point that the same trajectory 
can also be represented by parallel segments at angles
$\pi - \varphi$, $\pi + \varphi$ and $2\pi - \varphi$.
In general, the number of
directions for representing a trajectory equals the number of copies, $N$,   
that constitute the invariant surface.

\par The classical propagator on an invariant surface parametrised
by $\varphi$ is thus 

\be
L^t (\varphi) {\circ}\phi({\bf q}) = \int d{\bf q'}\;
\delta({\bf q} - {\bf q}'^t(\varphi))
\;\phi({\bf q}') \label{eq:prop2}
\ee

\noindent
where ${\bf q}$ refers to the position in the singly connected region 
and ${\bf q}'^t(\varphi)$ is the  
time evolution parametrised by $\varphi$ as described above.
We denote by $\{\Lambda_n(t;\varphi)\}$, the eigenvalues
of $L^t(\varphi)$ and from its multiplicative nature, it
follows that $\Lambda_n(t;\varphi) = e^{\lambda_n(\varphi)t}$.

\par The trace of $L^t$ takes into account all possible
invariant surfaces that exist and hence involves an additional integration
over $\varphi$ \cite{circle1}. Thus  

\be
{\rm Tr}~L^t = \int d\varphi~L^t(\varphi) = \int d\varphi \sum 
e^{\lambda_n(\varphi)t} =  
\int d\varphi \int d{\bf q} \;\delta({\bf q} -
{\bf q}^t(\varphi)) \label{eq:trace3}
\ee

\noindent
Clearly the only orbits that contribute are the ones that are
periodic. Further, the {\bf q} integrations are simpler if we
transform to a local coordinate system with one component
parallel to the trajectory and the other perpendicular.
Thus $\delta_{\|}(q_\| - q_\| ^t) = 
\sum_p \sum_r {1\over v}\delta (t-rT_p) $
where $v$ is the velocity, $T_p$ is the period of the
orbit and $r$ is the repetition number. Similarly, for
an orbit of period $rT_p$ parametrised by the angle
$\varphi_p$, $\delta_{\bot} (q_\bot - q_\bot ^{rT_p}) =
\delta (\varphi - \varphi_p)/{\left |\partial q_\bot/\partial
\varphi \right |_{\varphi = \varphi_p} }$ where 
$\left |\partial q_\bot/\partial \varphi 
\right |_{\varphi = \varphi_p} = rl_p$ for marginally
unstable billiards. Putting these results together and
noting that each periodic orbit occurs in general at
$N$ different values of $\varphi$,  we finally have
 
\be
{\rm Tr}~L^t =  N~\sum_p \sum_{r=1}^{\infty}
 {a_p\over rl_p}\delta(l-rl_p) \label{eq:trace4}
\ee

\noindent
where $l = tv$ and the summation over $p$ refers to all primitive 
periodic orbit families with length $l_p$ and  occupying an area $a_p$.
We thus have the analogue of Eq.~(\ref{eq:hbolic}) for integrable
and polygonal billiards.

\par In some cases, it is possible to interpret the periodic
orbit sum in Eq.~(\ref{eq:trace4}) starting with the appropriate
quantum trace formula \cite{neglect-isolated1} 

\be
\sum_n \delta (E - E_n) = d_{av}(E) + {1\over \sqrt{8\pi^3}}
\sum_p \sum_{r=1}^{\infty} {a_p\over \sqrt{krl_p}}\cos(krl_p -
 {\pi\over 4} - \pi rn_p - r\nu_p{\pi\over 2}) \label{eq:richens}
\ee

\noindent
Here $d_{av}(E)$ refers to the average
density of quantal eigenstates, $k = \sqrt{E}$, 
$l_p$ is the length of a primitive 
periodic orbit family, $n_p$ the number of bounces
that it suffers at the boundary and $\nu_p$ the number of 
caustics encountered by the orbit. Note that in the 
Neumann case, $n_p = 0$ since there is no phase loss 
on reflection while for 
polygonal billiards, $\nu_p = 0$. For convenience, 
we have set $\hbar = 1$ and the mass $m=1/2$.
Starting with the function  

\be
 \sum_n f(\sqrt{E_n}l) e^{-\beta E_n} =
\int_\epsilon^\infty dE\; f(\sqrt{E}l) e^{-\beta E} \sum_n \delta(E-E_n) 
\ee

\noindent
where $f(x) = \sqrt{{2\over \pi x}} \cos(x - \pi/4)$  
and $ 0 < \epsilon < E_0 $,
it is possible to show using Eq.~(\ref{eq:richens}) that
for polygonal billiards \cite{prl1} 

\be
\sum_p \sum_{r=1}^{\infty} {a_p (-1)^{rn_p} \over \sqrt{l}\sqrt{rl_p}}
 \delta (l-rl_p)
 = 2\pi b_0 + 2\pi \sum_n f(\sqrt{E_n}l) \label{eq:myown1}
\ee

\noindent 
for $\beta \rightarrow 0^+$. In the above, 
$b_0 = \sum_p \sum_r {a_p (-1)^{rn_p} \over 4\pi}\int_0^{\epsilon} dE
f(\sqrt{E}l)f(\sqrt{E}rl_p) $ and is a 
constant \cite{prl1,db1}.
It follows from Eqns.~(\ref{eq:myown1}) and~(\ref{eq:trace4})
and the fact that $h(x)\delta(x-a) = h(a)\delta(x-a)$ that  

\be
{\rm Tr}~L^t = 2\pi Nb_0 + 2\pi N \sum_n f(\sqrt{E_n}l) 
\label{eq:trace5}.  
\ee

\noindent
where $\{E_n\}$ are the Neumann eigenvalues of the system.
Thus $\lambda_n(\varphi) = i\sqrt{E_n}\sin(\varphi) \cite{asymp}$.
 
For integrable polygonal billiards, Eq.~(\ref{eq:trace5}) 
can in fact be derived directly from
the expression for $\Lambda_n(t)$ given earlier.
To illustrate this, we consider a rectangular billiard 
where the  
hamiltonian expressed in terms of the actions,
${I_1,I_2}$ is $H(I_1,I_2) = \pi^2(I_1^2/L_1^2 + I_2^2/L_2^2)$
where $L_1,L_2$ are the lengths of the two sides.
With $I_1 = \sqrt{E}L_1\cos(\varphi )/\pi$ and
$I_2 = \sqrt{E} L_2\sin(\varphi )/\pi$, it is easy to
see that at a given energy, $E$, each torus is parametrised by a
particular value of $\varphi$. Thus $\Lambda_{\bf n}(t;\varphi) =
e^{i2\pi t\sqrt{E}(n_1\cos(\varphi)/L_1 + n_2\sin(\varphi)/L_2)}$
and the spectrum is continuous, parametrized by $\varphi$.
The trace of the classical propagator is thus 

\be
{\rm Tr}~L^t = \sum_{\bf n} \int_{-\pi - \mu_n}^{\pi - \mu_n} d\varphi\;
e^{il\sqrt{E_{\bf n}}\sin(\varphi + \mu_{\bf n})}
= 2\pi \sum_{\bf n} J_0(\sqrt{E_{\bf n}}l)  \label{eq:bessel}
\ee

\noindent
where $l = 2t\sqrt{E}$, $\tan(\mu_{\bf n}) = n_1L_2/(n_2L_1)$
and $E_n = \pi^2(n_1^2/L_1^2 + n_2^2/L_2^2)$.
On separating out ${\bf n} = (0,0)$ from the rest, restricting
the summation to the first quadrant of the integer lattice and
noting that for a rectangle $b_0 = 1/4$, Eq.~(\ref{eq:trace5})
follows \cite{asymp,unit}.

\par The classical evolution operator thus serves to determine
the Neumann spectrum in polygonal billiards.
We have confirmed this numerically for the rectangular and
equilateral triangle billiards and an outline of the 
method used can be found towards the end of this paper.
Appropriate modifications however need to be made for the 
Dirichlet spectrum and for systems that have caustics
(the circle billiard is an example) and the construction 
of the evolution operator is then guided by the nature
of the quantum trace formula 

\par The {\it quasi-classical} evolution operator linking     
the classical and semiclassical eigenvalues can be defined as 

\be
L_{qc}^t(\varphi){\circ}\phi({\bf q}) = \int d{\bf q'}\;\delta({\bf q} - 
{\bf q}'^t(\varphi)) e^{-in(t)\pi - i\nu(t){\pi\over 2}}
\phi({\bf q}') \label{eq:prop3}
\ee

\noindent
where $\nu(t) = \nu({\bf q}^{-t}(\varphi))$
and $n(t) = n({\bf q}^{-t}(\varphi))$ count respectively 
the number of caustics and (phase altering)  
reflections encountered by the trajectory 
${\bf q}^{-t}(\varphi)$ in time $t$. 
As before, the multiplicative
nature of $L_{qc}^t(\varphi)$ implies that its spectrum 
is of the form $\{e^{\lambda_n(\varphi)t}\}$ and we shall now
show that for the quasi-classical operator defined in Eq.~(\ref{eq:prop3}),
$\{\lambda_n\}$ has a one-to-one correspondence with the 
appropriate quantum spectrum.

The trace of $L_{qc}$ is given by 

\be
{\rm Tr}~L_{qc}^t = \int_n d\varphi \sum e^{\lambda_n(\varphi)t} =  
\int d\varphi \int d{\bf q}\; \delta({\bf q} -
{\bf q}^t(\varphi))\; e^{-in(t)\pi - i\nu(t){\pi\over 2}} 
\label{eq:trace6}
\ee

\noindent
and the contributions to it are due to orbits that are
periodic. Since they occur in families, the
integrand is independent of ${\bf q}$ as shown
earlier and we have 

\bea
{\rm Tr}~L_{qc}^t & = & N~\sum_p \sum_{r=1}^{\infty} {a_p  \over rl_p}
\delta(l-rl_p) \int d\varphi \; \delta(\varphi - \varphi_p)
e^{-in({\bf q}^{-t}(\varphi))\pi -
i\nu({\bf q}^{-t}(\varphi)){\pi\over 2}} \\
& = & N~\sum_p \sum_{r=1}^{\infty} {a_p  \over rl_p}
\delta(l-rl_p) e^{-irn_p\pi - ir\nu_p{\pi\over 2}} \label{eq:trace7}
\eea

\noindent
Starting with the function $\sum_n\; g(\sqrt{E_n}l)\exp(-\beta E_n)$, it
follows from Eq.~(\ref{eq:richens}) that for $\beta \rightarrow 0^+$, 

\be
{\rm Tr}~L_{qc}^t = \sum_n \Lambda_n(t) = 
2\pi N C + 2\pi N \sum_n g(\sqrt{E_n}l) \label{eq:pre_final}
\ee

\noindent
where $\{E_n\}$ now refers to the desired quantum spectrum,
$g(x) = \sqrt{2\over \pi x} {\rm exp}(-ix + \pi/4)$ and $C$ is a constant. 
It follows from 

\be
g(x) \simeq {1\over 2\pi}\int_0^{2\pi}~e^{-i\sqrt{E_n}l\sin(\varphi/2)}d\varphi
\ee

\noindent
that 

\be 
\lambda_n(\varphi)= i\sqrt{E_n}\sin(\varphi/2) \label{eq:final}
\ee
 
\noindent
Eq.~(\ref{eq:final}) forms the central result of this paper. 

\par In order to numerically demonstrate our results, we first
note that 

\be
L^t_{qc}(\varphi) \circ \phi({\bf q}) = \phi(q^{-t}(\varphi))
e^{-in(q^{-t}(\varphi))\pi - i\nu(q^{-t}(\varphi)){\pi\over 2}}
= \sum_n c_n e^{i\sqrt{E_n}\sin(\varphi/2)l} \phi_n(q) \label{eq:expand}
\ee 

\noindent 
where $\{\phi_n\}$ are the eigenfunctions of $L^t_{qc}$ and
$\{c_n\}$ are the expansion coefficients for the function
$\phi({\bf q})$. Thus on evolving $\psi(t;\varphi) = 
\phi(q^{t}(\varphi))e^{-in(q^{t}(\varphi))\pi - 
i\nu(q^{t}(\varphi)){\pi\over 2}}$ and averaging this over
$\varphi$, we can extract the quantum eigenvalues
from the power spectrum of $\langle \psi(t) \rangle$
where $\langle \ldots \rangle$ denotes averaging over $\varphi$.
In practice though, we use 
the function $\langle \psi(t) \rangle e^{-\beta t^2}$ to compensate  
for the finite data set with $\beta > 0$.

\par The first application of the quasi-classical evolution
operator is provided for a circle billiard of unit radius
in Fig.~(2). The function $\phi({\bf q})$ is chosen to be
a Gaussian inside a cell centred at ${\bf q}_c$ and $0$ outside
\cite{discont}. The power spectrum, $S(k)$ is displayed
in Fig.~(2) where the arrows mark the positions of the Dirichlet
EBK eigenvalues \cite{diff}. 

\par
\vbox{
\begin{figure}[hbt]
\centerline{\hspace{0.85in}
\epsfig{figure=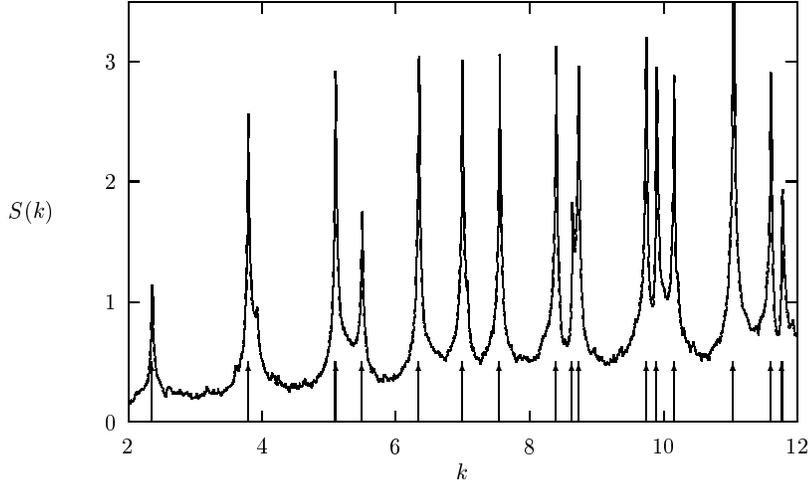,height=3in}}
\caption{Power spectrum of $\langle \psi(t) \rangle$
for a circle billiard of unit radius. The arrows mark
the position of the Dirichlet EBK eigenvalues. }
\end{figure}
}

\noindent
A total of 1000 arbitrary trajectories with 
different values of angular momentum have been used and the
the power spectrum is averaged over 4 cells.   
We have also generated the Neumann spectrum of the circle
billiard (this is distinct from the Dirichlet spectrum) 
by setting $n(t) = 0$.

\par Finally, we provide results for a non-integrable triangular 
billiard with base angles ($\pi/5,3\pi/10$) in Fig.~(3). 
The function $\langle \psi(t) \rangle$
has been generated using 300 arbitrary trajectories 
and averaging over 150 cells while  
the power spectrum has been obtained with $\beta = 0.006$.
We find convergence to be rapid with the number
of trajectories used. The width of individual peaks depends
on $\sqrt{\beta}$  and the choice of $\beta$ is dictated by
several factors which include
the degree of smoothening attained by averaging over different
cells and the length of trajectories. In the present context,
the value of $\beta$ is also influenced by the height of the
shortest peak (at 18.5) and is chosen such that it is distinct
from the noisy baseline. As in the circle billiard \cite{diff}, peak positions
in the non-integrable case
do not coincide with the exact quantum spectrum.
The average
error is about 8 percent of the mean level spacing
in the triangular billiard and we find this to be of the
same magnitude as in the eigenvalues obtained from
Bogomolny's \cite{bogo} transfer operator ( see Fig.~(3) ).

\par
\vbox{
\begin{figure}[hbt]
\centerline{\hspace{0.85in}
\epsfig{figure=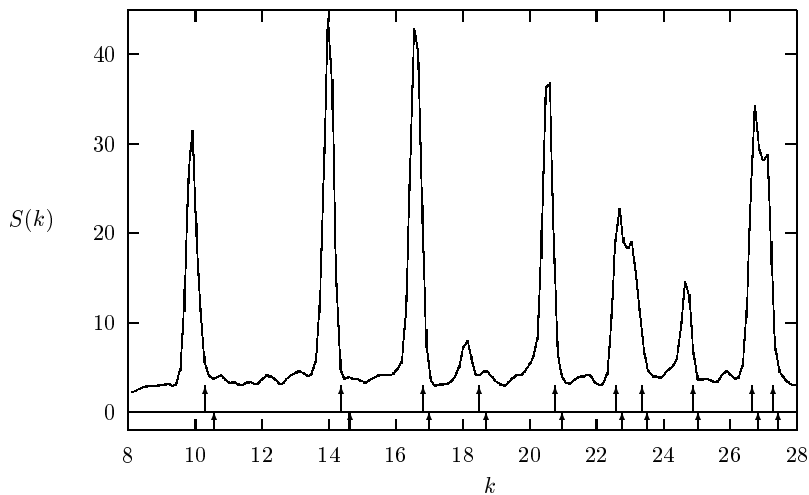,height=3in}}
\caption{Power spectrum of $\langle \psi(t) \rangle$
for a non-integrable triangular
billiard with base angles ($\pi/5$,$3\pi/10$). The upper set of arrows
mark the positions of the {\it exact} Dirichlet eigenvalues while
the lower set is obtained from Bogomolny's transfer operator.}
\end{figure}
}

\noindent
The neglect
of higher order corrections thus introduces errors as in 
all semiclassical schemes and it is not immediately obvious 
how effects such as diffraction can be taken into account
in the quantization recipe developed here. We wish to point 
out however that direct applications of periodic orbits
have met with little success in determining individual quantum
levels of pseudointegrable systems \cite{PJ2} 
and the results provided here are thus of some significance.

\par In addition to the cases presented here, we have also
generated the semiclassical spectrum of a circle billiard 
with a single magnetic flux line passing through the centre as
well as other non-integrable triangular billiards using
arbitrary trajectories \cite{dbprep}. 
Also, it is possible to generate excited states by
suitably choosing the function $\phi({\bf q})$ such that
it has nodes. We have thus been able to generate the 
first 100 levels of an arbitrary triangle \cite{dbprep}.

\par We provide below a summary of our results : 
\vskip 0.1 in
$\bullet$ (i)~we have introduced the classical evolution
operator for marginally unstable systems and shown 
that for polygonal billiards, the classical spectrum
on an invariant surface labelled by $\varphi$ is
$\{e^{i\sqrt{E_n}\sin(\varphi)t}\}$ where $\{E_n\}$ is
the (quantum) Neumann spectrum.

$\bullet$ (ii)~we have also shown how the classical evolution
operator 
can be appropriately modified to construct a quasi-classical
adaptation whose spectrum has a one-to-one correspondence
with the desired semiclassical
quantum spectrum ; 

$\bullet$ this yields a quantization procedure using
arbitrary trajectories and its utility is demonstrated by
determining individual levels in a class of systems where
direct applications of periodic orbits have met with little
success so far.

\par It is a pleasure to acknowledge several useful discussions with
Predrag Cvitanovi\'{c} and Gregor Tanner.

\tighten


\begin{references}

\bibitem{PC} P.Cvitanovi\'{c}~et.~al.,~{\it Classical~and~Quantum~Chaos~for~Cyclists},
http://cats.nbi.dk. 
\bibitem{MC} M.C.Gutzwiller, {\it Chaos in Classical and Quantum Mechanics},
Springer, New York, 1990.
\bibitem{PCBE} P.Cvitanovi\'{c} and B.Eckhardt,
J. Phys. A. {\bf 24}, L237 (1991).
\bibitem{notL2} We restrict here to the space of analytic functions. In 
the space of square integrable functions, $L^t$ is unitary \cite{LM-book}.
\bibitem{LM-book} A.Lasota and M.Mackey, {\it Chaos, Fractals, and Noise},
Springer-Verlag, New York, 1994. 
\bibitem{exception} For some exceptional cases like the motion on a
compact surface of constant negative curvature, there exists
a one-to-one correspondence between $\{\lambda_n\}$ and
$\{E_n\}$ \cite{prl3}.
\bibitem{prl3} D.Biswas and S.Sinha,  Phys. Rev. Lett.,
{\bf 71}, 3790(1993).
\bibitem{vattay1} P.Cvitanovi\'{c} and G.Vattay, Phys. Rev. Lett.,
{\bf 71}, 4138(1993).
\bibitem{heller} E.J.Heller and S.Tomsovic, Physics Today, 38-46, July 1993
and references therein. 
\bibitem{dbprep} Details of this work together with a study of
systems other than billiards will be published elsewhere.
\bibitem{PJ} P.J.Richens and M.V.Berry, Physica D{\bf 2}, 495(1981).
\bibitem{keller} J.B.Keller and S.I.Rubinow, Ann. Phys. (N.Y.)
{\bf 9}, 24(1960).
\bibitem{circle1} For the circle billiard, the angular momentum
is conserved and  $\varphi$ is then a measure of the impact
parameter, $R_0 = R \sin(\varphi)$ where $R$ is the radius
of the circle.
\bibitem{neglect-isolated} We have neglected here the contribution
to the trace from isolated periodic orbits. As in the quantum
case \cite{neglect-isolated1}, we expect them to be less important.
\bibitem{neglect-isolated1} The respective contributions
of isolated and diffractive periodic orbits are
$O(k^{-1})$ and $O(k^{-1-\nu/2})$ where $\nu$ is the total
number of (singular) vertex connections in a diffractive
periodic orbit.
\bibitem{prl1} D.Biswas and S.Sinha, Phys. Rev. Lett.
 {\bf 70}, 916 (1993).
\bibitem{db1}  Numerical verification of this for a generic polygon 
can be found in D.Biswas, Phys. Rev. {\bf E54}, R1044 (1996).
\bibitem{asymp} The function $f(x)$ is the asymptotic expansion of
$J_0(x)$.
\bibitem{unit} Note that the term $2\pi b_0 N$ in Eq.~(\ref{eq:trace5}) 
reflects the fact that there is a unit eigenvalue at each $\varphi$.
\bibitem{discont} The results are better
for functions $\psi({\bf q})$ that are discontinuous possibly because
all modes are excited. We have obtained the same results with
function other than a Gaussian. In particular, $\psi({\bf q})$
can be defined to be $1$ inside the cell and $0$ outside and
$\langle \psi(t) \rangle$ is then a weighted recurrence fraction
of trajectories.
\bibitem{diff} The exact quantum eigenvalues are different from
the EBK eigenvalues for the circle billiard \cite{keller}.
\bibitem{bogo} E.B.Bogomolny, Nonlinearity {\bf 5}, 805(1992). 
\bibitem{PJ2} P.J.Richens, J.Phys {\bf A 16}, 3961(1983); 
Y.Shimizu and A.Shudo, Chaos Solitons and Fractals,
{\bf 5}, 1337(1995).

\end{references}
\end{document}